\documentstyle[preprint,prb,aps,epsfig]{revtex}
\begin{document}
\draft
\title{Sensitivity of a micromechanical displacement detector
based on the radio-frequency single-electron transistor}

\author{Miles P. Blencowe and Martin N. Wybourne}
\address{Department of Physics and Astronomy, Dartmouth College, 
Hanover, New Hampshire 03755}
\date{\today}
\maketitle
\begin{abstract}
We investigate the tunneling shot noise limits on the sensitivity of a  
micromechanical displacement detector based on a metal junction, radio-frequency 
single-electron transistor (rf-SET). In contrast with the charge 
sensitivity of the rf-SET electrometer, the displacement sensitivity improves 
with 
increasing gate voltage bias and, with a suitably optimized rf-SET, 
displacement sensitivities of $10^{-6}~{\rm\AA}/\sqrt{\rm Hz}$ 
may be possible.
\end{abstract}


Recent advances in microfabrication technology have lead to renewed 
interest in demonstrating quantum behavior in small mechanical 
systems. Proposals include quantum tunneling,\cite{cleland1} quantum 
superposition states,\cite{bose} and the generation of nonclassical, 
squeezed states.\cite{blencowe} A common requirement  is the need to
resolve sub-{\AA}ngstrom displacements at radio 
frequencies. Considering, for example,  
a  micron-sized 
cantilever with a realisable resonant frequency of 100~MHz and a
quality factor $Q\sim 10^{4}$,\cite{cleland1,cleland2} a displacement 
sensitivity of 
about $10^{-4}$~{\AA} is required to detect quantum squeezing.\cite{blencowe} 
Two methods for 
detecting motion of (sub)micron-scale structures are the magnetomotive 
technique\cite{mohanty} and the optical interferometric technique,\cite{carr}
where  
sensitivities before signal averaging
of around $10^{-2}~{\rm\AA}/\sqrt{\rm Hz}$ at 1~MHz 
 were reported.

In the present letter, we consider the possibility of a metal junction 
single-electron transistor (SET)-based  displacement detector for 
resolving sub-{\AA}ngstrom 
motion of micron-scale and smaller mechanical resonators with 
fundamental frequencies  up to a few hundred MHz.
 A radio-frequency single-electron transistor (rf-SET) was developed a 
few years ago\cite{schoelkopf} to overcome 
the limited bandwidth available 
in conventional 
SETs, and a rf-SET was recently used to investigate charge noise in  
an AlGaAs/GaAs 
quantum dot.\cite{fujisawa} In the original device,\cite{schoelkopf}
a bandwidth of around 100~MHz was demonstrated, a 
considerable improvement over conventional SET bandwidths of typically
only a few kHz. The charge sensitivity of the rf-SET was 
$1.2 \times 10^{-5} e/\sqrt{\rm{Hz}}$ 
at $1.1$~MHz, comparable to the best sensitivities of 
low frequency SETs.\cite{starmark} 
In a recent theoretical analysis,\cite{korotkov} 
Korotkov and Paalanen established that an additional order of 
magnitude improvement in the charge sensitivity could be expected 
with a suitably optimized device. 

With the ability to operate at 
signal frequencies in the radio frequency range, we can expect considerable 
improvements in $1/f$ noise which limits the sensitivity of 
conventional SETs.
Taking as our starting point the  noise analysis of Korotkov 
{\it et al.},\cite{korotkov} 
we show below that metal SET  displacement sensors may achieve 
shot-noise limited sensitivities   
of $10^{-6}~{\rm\AA}/\sqrt{\rm Hz}$. For the above 
cantilever example, this corresponds 
to an absolute displacement sensitivity of $10^{-4}$~{\AA} at 100 
MHz, adequate for detecting quantum-squeezed motion. 

With such improvements in displacement sensitivity, 
significant advances could also be made in   understanding how mechanical 
energy is dissipated in micromechanical resonators. This problem is 
motivated in part by the increasing use of (sub)micron-sized 
cantilevers as sensors in new forms of 
force microscopy, such as the magnetic resonance force microscope, 
where the need to detect extremely weak forces requires smaller mass 
cantilevers with longer mechanical damping times.\cite{sidles}       
Recent experiments which measure $Q$ 
have provided strong evidence for surface defect relaxation  as the 
dominant mechanism limiting the quality factor.\cite{cleland2,carr,yasumura} 
However,  more 
information could be gained about the relaxation dynamics if
the damping motion of a micromechanical 
resonator could be measured with sufficient time and displacement 
accuracy to resolve 
the   mechanical, random telegraph signals due to {\it single} 
defects undergoing relaxation.

The schema of the rf-SET displacement detector is shown in Fig.~1.
The basic principle of the 
device involves locating one of the  gate 
capacitor plates of the SET  on the cantilever so that, 
for  fixed gate voltage bias,  a mechanical 
displacement is converted into a polarization charge fluctuation. 
The  
stray capacitance $C_{\rm s}$ of the leads contacting the SET and an inductor 
 $L$ form a  tank circuit with resonant  frequency 
$\omega_{\rm T}=(LC_{\rm s})^{-1/2}$, loaded by the SET. A 
monochromatic carrier wave is sent down the cable. At 
the resonant frequency the circuit impedance is small and the 
reflected power provides a measure of the SET's differential 
resistance $R_{d}$.\cite{korotkov} When the gate capacitor is biased,
mechanical motion of the cantilever is converted into differential 
resistance changes, hence modulating the reflected signal power.

Consider an incoming wave of the form 
$V_{\rm in}\cos \omega t$ at the  end of the cable. The reflected wave 
is $V_{\rm ref}(t)= v(t)-V_{\rm in}\cos \omega t$, where the  
voltage $v(t)$ at the end of the cable satisfies the differential equation
\begin{equation}
    \ddot{v} L C_{s} +\dot{v} R_{0} C_{s} +v=2 (1-\omega^{2} L C_{s}) 
    V_{\rm in}\cos \omega t -R_{0} I_{SD}(t), 
    \label{diffeq}
\end{equation}
with $R_{0}$ the impedance of the cable and $I_{SD}(t)$ the SET 
source-drain current. Setting $\omega=\omega_{\rm T}$
and substituting the Fourier 
decomposition $v(t)=\sum_{n=1}^{\infty}(X_{n}\cos n\omega t + Y_{n}\sin 
n\omega t)$ into Eq. (\ref{diffeq}),   we obtain for the Fourier coefficients
\begin{eqnarray}
X_{1}&=&2\sqrt{L/C_{s}}\langle I_{SD}(t) \sin\omega 
t\rangle,\label{x1harmonic}\\
Y_{1}&=&-2\sqrt{L/C_{s}}\langle I_{SD}(t) \cos\omega 
t\rangle,\label{y1harmonic}
\end{eqnarray}
where $\langle\cdots\rangle$ denotes the time  average and we restrict 
the analysis 
to the first harmonic  of the reflected wave.  The  current 
$I_{SD}(t)$ depends on the voltage $V_{SD}(t)$ across the SET, which 
in turn depends on $v(t)$: 
\begin{equation}
    V_{SD}(t)=L R_{0}^{-1}[2 V_{\rm in} \omega\sin\omega t+\dot{v}(t)] +v(t). 
\label{vsd}
\end{equation}
The Fourier coefficients $X_{1}$, $Y_{1}$ are found by solving iteratively 
Eqs. (\ref{x1harmonic}) and (\ref{y1harmonic}). However, in the regime
 $R_{d} \sqrt{C_{s}/L}\gg Q_{\rm T}\gg 1$, where the tank circuit 
 quality factor  $Q_{\rm 
T}=R_{0}^{-1}\sqrt{L/C_{s}}$, 
we can approximate Eq. (\ref{vsd}) as 
$V_{SD}(t)=2 Q_{\rm T} V_{\rm in}\sin\omega t$,  thus the 
coefficients $X_{1}$, $Y_{1}$ are approximately solved for once the $I_{SD}(t)$
dependence on $V_{SD}(t)$ is known. 
We will also assume that the carrier frequency satisfies 
$\omega\ll \sqrt{\langle 
I_{SD}^{2}\rangle}/e$. Under these conditions the current $I_{SD}(t)$ maintains a fixed phase 
relationship with respect to
$V_{SD}(t)$ and, from the  approximate form of 
$V_{SD}(t)$ and Eq. (\ref{y1harmonic}), we see that the 
time-averaging gives $Y_{1}=0$.

At the mechanical signal frequency $f_{s}$, where $2\pi f_{s}\lesssim 
\omega_{\rm T}/Q_{\rm T}$  (for example, with $\omega_{\rm 
T}=2\pi\times 1$~GHz and 
$Q_{\rm T}=10$, we have $f_{s}\lesssim$~100~MHz), 
the minimum detectable displacement 
$\delta x$ 
in terms of the spectral density $S_{X}(f_{s})$ of fluctuations in 
$X_{1}(t)$  is
\begin{equation}
    \delta x =\sqrt{S_{X}(f_{s})\Delta f}/|dX_{1}/dx|,
    \label{sensitivity1}
\end{equation}
where $\Delta f$ is the signal bandwidth and $x$ denotes the 
displacement from equilibrium separation of the gate capacitor plates. 
The  fundamental 
limit on $\delta x$ is given by the intrinsic shot 
noise due to 
the SET tunneling current. Using the shot noise formula $S_{I}=2 e I$, 
which is
approximately valid close to the tunneling threshold, and using Eq. 
(\ref{x1harmonic}) to relate $S_{X}$ to the current spectral density 
$S_{I}$, Eq. (\ref{sensitivity1}) becomes
\begin{equation}
    \delta x =\sqrt{2 e \langle |I_{SD}(t)|\sin^{2}\omega 
    t\rangle\Delta f}/|\langle dI_{SD}(t)/dx\ \sin\omega t\rangle|.
    \label{sensitivity2}
\end{equation}

All that remains is to determine the $I_{SD}(t)$ dependence on 
$V_{SD}(t)$ and substitute into Eq. (\ref{sensitivity2}). 
We use the ``orthodox'' 
theory\cite{averin} and the method of analytic solution given in 
Ref.\ \onlinecite{amman}.  Refering to Fig.~1, and as usual defining 
$C_{\Sigma}=C_{1}+C_{2}+C_{g}$,  when the  voltage amplitude across 
the SET $A=2 Q_{\rm T} 
V_{\rm in}$ is small compared to the voltage $e/C_{\Sigma}$, and also the 
thermal energy $k_{B} T\ll e A$, 
 then the current in the tunneling region between stable 
 regions of $n$ and $n+1$ 
excess electrons on the SET island can be 
well-approximated as
\begin{eqnarray}
    I_{SD}&=& 
    e[b_{1}(n)-t_{1}(n)]\rho(n)+e[b_{1}(n+1)-t_{1}(n+1)]\rho(n+1)\cr
    &=&e[b_{2}(n)-t_{2}(n)]\rho(n)+e[b_{2}(n+1)-t_{2}(n+1)]\rho(n+1).
\end{eqnarray}
Here the probabilities $\rho(n)$, $\rho(n+1)$  are given approximately as:
\begin{eqnarray}
    \rho(n)&=&[t_{1}(n+1)+b_{2}(n+1)]/[b_{1}(n)+t_{2}(n)+t_{1}(n+1)+
    b_{2}(n+1)],\cr
    \rho(n+1)&=&[b_{1}(n)+t_{2}(n)]/[b_{1}(n)+t_{2}(n)+t_{1}(n+1)+b_{2}(n+1)].
    \label{rhon}
\end{eqnarray}
In this approximation 
the  tunnel current peaks are well-separated in gate voltage $V_{g}$. 
The  tunneling rates $b_{i}\ (t_{i})$ from the bottom 
(top) across the $i$th junction of the SET take
the usual form.\cite{details}
 
In Fig.~2 we show the dependence of the minimum detectable mechanical 
displacement
$\delta x$ on the gate voltage $V_{g}$, ranging over the first few 
current peaks. These results assume 
 junction capacitance and resistance values 
$C_{1}=C_{2}=0.25~{\rm fF}$ and  $R_{1}=R_{2}=50~{\rm k}\Omega$, and 
a static gate capacitance $C_{g}=0.1~{\rm fF}$. We have also explored the 
dependence of $\delta x$ (optimized over 
$V_{g}$ for given $n$) on the SET voltage amplitude $A$ and also on an 
applied dc-voltage component. As was found for the charge sensitivity 
analysis,\cite{korotkov} further improvements in the optimized $\delta x$
are small.
  
Table~I shows a sampling of  $\delta x$ values
optimized with respect to $V_{g}$ 
over a range corresponding to the rising (left) side of a given 
current peak. 
Importantly, the optimized minimum displacement resolution  
improves 
with increasing current peak number $n$. This is in marked contrast 
to the optimized minimum detectable charge which is  
 independent of peak number.
This trend begs the question of how 
large a 
gate voltage can be applied to a metal junction SET. 
Given that 
electrometry is the 
most often considered SET application, for which
there is no gain in charge sensitivity with 
increasing $V_{g}$ (one  current peak is as good as any other),
this question  has apparently received little 
attention. For displacement detection on the other hand, 
the optimized sensitivity 
improves with increasing $V_{g}$.
There are no obvious reasons why the 
periodic current oscillations for a metal SET 
should not survive   all the way up to the breakdown voltage between 
the gate capacitor plates. 
As  $V_{g}$ increases, progressively more electrons 
tunnel onto the island, screening the increasing polarization charge so that 
the periodicity in the $V_{g}$ dependences of
source-island and island-drain junction 
voltages is maintained. 
Assuming the gate capacitor plates are 
separated by a vacuum, and 
taking a typical vacuum breakdown voltage  of $10^{8}$~V/m,\cite{ma} this 
suggests a maximum $V_{g}$ of around 10 V across a $0.1~\mu{\rm m}$ gap, on 
the  order of the upper limit considered in the above noise analysis 
(Table I).

An alternative scenario to the one just outlined is suggested by the fact 
that fluctuators in the vicinity of the SET island can result in a 
strong gate-voltage dependence for the $1/f$ 
source-drain current noise.\cite{zimmerman,smith} 
This raises the  possibility of  an 
optimum gate bias beyond which a further improvement in 
displacement sensitivity as predicted by the above shot noise analysis 
is offset by an increase in $1/f$ noise. Induced stresses in the gated 
cantilever as a result of its flexing motion may also affect the $1/f$ 
noise levels.\cite{martinis}

To conclude, we have obtained the fundamental, shot noise limiting 
sensitivity of an rf-SET based displacement detector. In contrast 
with the charge sensitivity of an rf-SET electrometer, 
the optimized displacement sensitivity 
improves with increasing gate voltage.  
Sensitivities of $10^{-6}~{\rm\AA}/\sqrt{\rm Hz}$ 
may be possible, depending on one or other of the  
$1/f$ noise gate voltage dependence  or  the gate capacitance breakdown 
characteristics.

\acknowledgements We thank Chris Berven for helpful conversations.

\begin{table}
\caption{Minimum detectable displacement $\delta x$ optimized with respect to 
$V_{g}$ about a sampling of current peaks for increasing $n$. Note the 
linear dependence of the optimized $\delta x$  on $V_{g}$  for $n\gg 1$.}
\label{Table 1}
\begin{tabular}{| l | l | l |}\hline
$n$ & $V_{g}~({\rm Volts})$ & $\delta x ({\rm \AA})/\sqrt{\rm Hz}$\\
\hline
$1$&$2.5\times 10^{-3}$&$2.2\times 10^{-3}$\\
$10$&$1.9\times 10^{-2}$&$2.9\times 10^{-4}$\\
$10^{2}$&$0.18$&$3.0\times 10^{-5}$\\
$10^{3}$&$1.8$&$3.0\times 10^{-6}$\\
$10^{4}$&$18.1$&$3.0\times 10^{-7}$\\
\hline
\end{tabular}
\end{table}

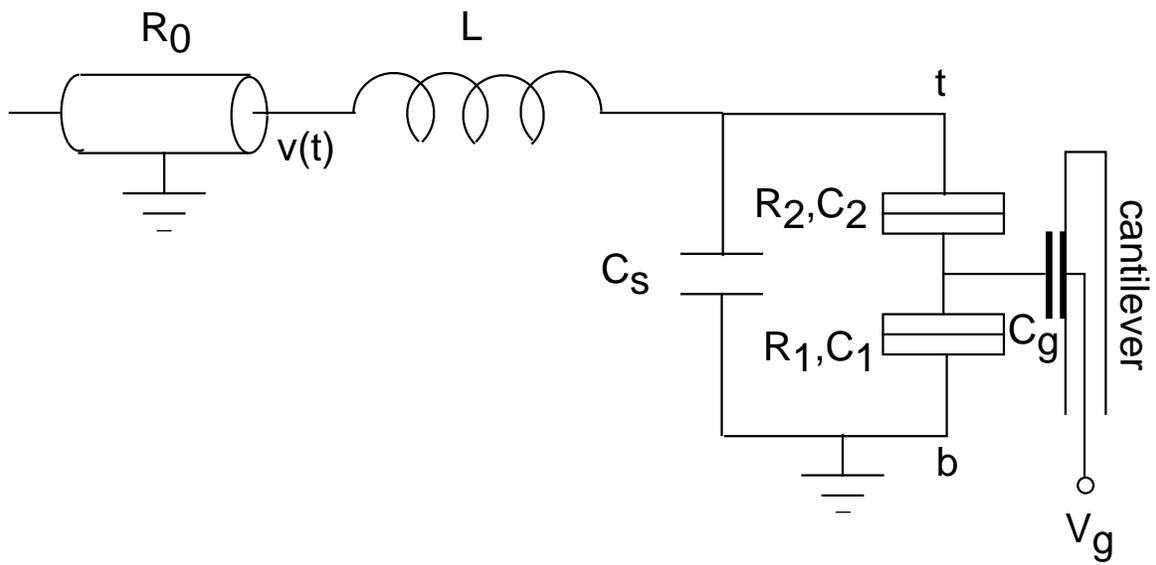
\begin{figure}
\caption{Schema of the rf-SET displacement detector.}
\label{Fig.1}
\end{figure}
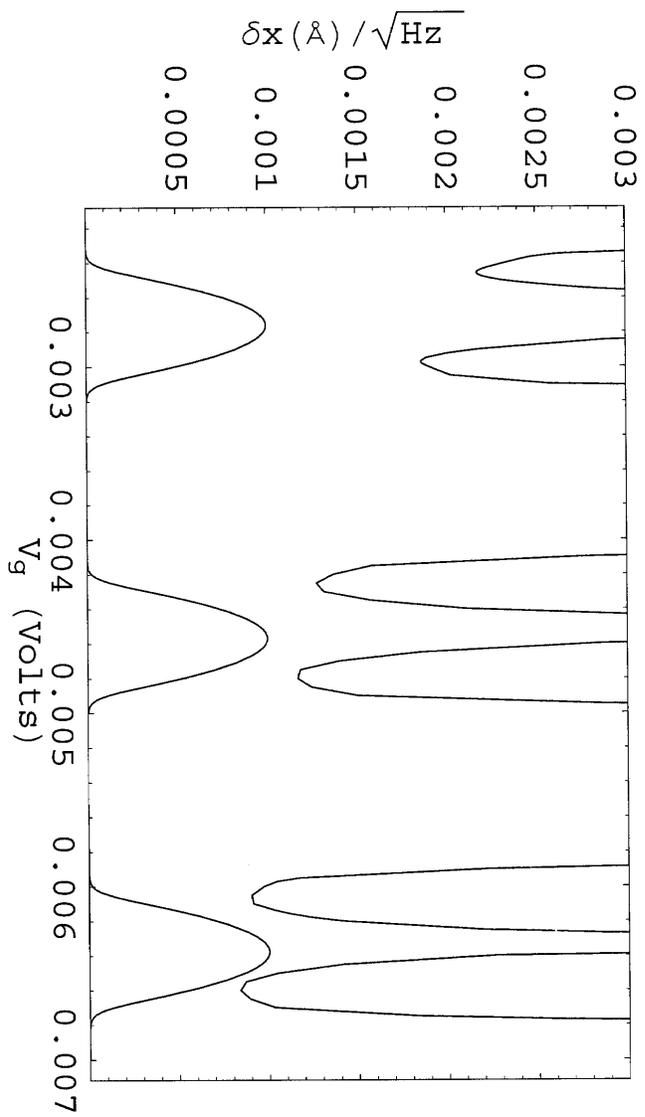
\begin{figure}
\caption{Minimum detectable displacement as a function of gate voltage 
(upper, `doublet' curves) 
corresponding to the first three $I_{SD}$ current amplitude peaks. The 
current amplitude versus gate voltage is also shown (lower curve - 
arbitrary scale)
for reference. The noise analysis assumes a   
 symmetric rf-SET at $T=30$~mK ($=0.01~e^{2}/C_{\Sigma}$) with 
 junction capacitances $C_{1}=C_{2}=0.25~{\rm fF}$, junction 
resistances $R_{1}=R_{2}=50~{\rm k}\Omega$, and static gate capacitance 
$0.1~{\rm fF}$ 
(corresponding to $1~\mu{\rm m}^{2}$ plate area and $0.1~\mu{\rm m}$ 
plate gap). The 
source-drain rf bias voltage amplitude is $A=10^{-4}~{\rm V}\ (=0.4~
e/C_{\Sigma})$. }
\label{Fig.2}
\end{figure}
\vfill
\eject

\mbox{\epsfig{file=rf-SETschema.EPSF, width=6in}}

\mbox{\epsfig{file=noiseplot2.EPSF, width=5.5in}}

\end{document}